\documentclass[two-column, nofootinbib]{revtex4-1}

\usepackage{amsmath}
\usepackage{tabularx}
\usepackage{graphicx}
\usepackage[usenames,dvipsnames,svgnames]{xcolor}
\usepackage{hyperref}

\hypersetup{dvips,dvipdfm,colorlinks=true,urlcolor=blue,filecolor=magenta,linktoc=page,citecolor=red,linkcolor=blue,bookmarks=true}

\begin{document}

\title{Testing Lambert $W$ equation of state with observational Hubble parameter data}
\author{Abdulla Al Mamon\footnote {Electronic Address: \texttt{\color{blue} abdulla.physics@gmail.com}}}
\affiliation{Department of Physics, \\ Vivekananda Satavarshiki Mahavidyalaya (affiliated
to the Vidyasagar University), Manikpara-721513, West Bengal, India.}

\author{Subhajit Saha\footnote {Electronic Address: \texttt{\color{blue} subhajit1729@gmail.com}}}
\affiliation{Department of Mathematics, \\ Panihati Mahavidyalaya, Kolkata 700110, West Bengal, India.}

\begin{abstract}
In this paper, we investigate the possibility that the Universe is driven by a single dark fluid described by a Lambert $W$ equation of state parameter, $w_{eff}$, which is essentially dependent on two parameters $\vartheta_{1}$ and $\vartheta_{2}$ which need to be fixed from observations. We obtain the constraints on these parameters using the latest 51 data points of $H(z)$ measurements, spanning the redshift range $0.07\leq z \leq 2.36$. The present study shows that the Universe is indeed undergoing an accelerated expansion phase following the decelerated one at the transition redshift, $z_{t}=0.77\pm0.03$ ($1\sigma$) and is well consistent with the recent observations. We also find that at low redshifts, $w_{eff}$ evolves only in the quintessence regime ($-1<w_{eff}<-\frac{1}{3}$) within $1\sigma$ confidence level. Its present value is found to be $-0.96\pm0.02$ ($1\sigma$). The fact that the present value of $w_{eff}$ is very close to the Cosmological Constant $\Lambda$ implies that our proposed equation of state parameter might serve as a unification of dark matter and dark energy. Furthermore, we compare the evolution of $H(z)$ for the model under consideration with that of the $\Lambda$CDM model. Finally, we observe that for the best-fit case, the differences between the two models are negligible at $z\sim 0.67$.
\begin{center}
\end{center}

\end{abstract}

\maketitle
\section{Introduction}
In 1998, two independent teams of cosmologists, the High-Z Supernova Search Team founded by B.P. Schmidt \cite{Schmidt1} and led by A. Riess et al. \cite{Riess1} and the Supernova Cosmology Project led by S. Perlmutter et al. \cite{Perlmutter1} analyzed observational evidences from Supernovae Type Ia (SNIa), cosmic microwave background (CMB) radiation, baryon acoustic oscillations (BAO), large scale structure (LSS) of spacetime, and weak lensing and established that our Universe is presently exhibiting a phase of accelerated expansion. The whole astronomical community was startled at this discovery because contemporary theoretical Cosmology had predicted a decelerated, matter dominated universe. Due to this unexpected result, cosmologists were forced to modify the standard model of Cosmology so that this new observational result could be incorporated into the theory. To this effect, most cosmologists took either of the following two ways---
\begin{enumerate}
\item [(a)] The domain of the stress-energy tensor, $T_{\mu\nu}$, was extended to include a dark energy component, a fluid with exotic properties such as a huge negative pressure. These type of fluid models later came to be known as modified matter models.
\item [(b)] The geometric part of the Einstein's field equations was modified to obtain a gravity theory different from General Relativity. These type of models later came to be known as modified gravity models.
\end{enumerate}
For an extensive review on the two approaches, one may see Refs. \cite{Copeland1,Frieman1,Nojiri1,Capozziello1,Nojiri2,Capozziello2,Matarrese1}. One must note that the the $\Lambda$-Cold-Dark-Matter ($\Lambda$CDM) model, regarded as the simplest modified matter model, is identified as the standard model in 21st century cosmology. This model consists of a tiny cosmological constant ($\Lambda$) which acts as {\it dark energy} (the dominant component) and {\it cold dark matter} in the form of dust. These two entities together make up almost 96\% of the energy budget of the Universe. However, the cosmological constant is plagued with several problems, particularly, the {\it fine tuning} and the cosmic {\it coincidence} problems. In addition to this, recent local measurement of Hubble constant $H_{0}$ by Hubble Space Telescope \cite{r19} is in disagreement with the $\Lambda$CDM cosmology. These have prompted cosmologists to devise alternative dark energy models with the assumption that the cosmological constant problem is solved in such a way that $\Lambda$ vanishes completely. On the other hand, dark energy is usually characterized by an dynamical effective {\it equation of state} (EoS) parameter. A large number of functional forms for the EoS parameter have been studied to account for this unknown component. For reviews on the various dark energy candidates, one can refer to \cite{Copeland1,Frieman1,Matarrese1}. The proposed candidates for the EoS parameter for dark energy are constrained with different observational data sets in order to check the viability of a particular model. Although these models fit the observational data sets quite well, yet all of these models have their own demerits. Hence the study of cosmic acceleration will continue for the foreseeable future.\\ 
\par In this context, very recently Saha and Bamba \cite{Saha1} have introduced a new fluid which deals with a special mathematical function, known as the Lambert $W$ function. The functional form of the proposed Lambert $W$ EoS parameter is not very simple and straight forward. As we shall see, the EoS parameter of this fluid has two free parameters, $\vartheta_1$ and $\vartheta_2$. Using some motivated choices of these parameters, they have shown that this new fluid can, in principle, explain the evolutionary stages of the Universe. This fact has motivated us to study the model more deeply in order to have a better understanding of the effect of this special function in the observational point of view. In the present work, we wish to constrain these free parameters ($\vartheta_1$ and $\vartheta_2$) using the latest 51 data points of $H(z)$ measurements, spanning the redshift range $0.07\leq z \leq 2.36$. We would like to emphasize that this work represents the first observational study on this new fluid. Using the best fit values of $\vartheta _\mathrm{1}$ and $\vartheta _\mathrm{2}$, we then reconstruct the evolutions of the effective EoS parameter and the deceleration parameter for the Lambert $W$ model. We also study the evolutions of the Hubble parameter and the distance modulus for the present  model and the standard $\Lambda$CDM model and compare that with the observational datasets.\\
\par The structure of this paper is as follows. In section \ref{sec-lambert}, we introduce the basic properties of the Lambert $W$ function. We briefly describe the Lambert $W$ cosmological model in section \ref{sec-model}. In section \ref{sec-obs}, we discuss the dataset and method used in this work along with the results obtained from the analysis of observational data. Finally, in section \ref{sec-con}, we present our conclusions. \\
\par Throughout the text, the symbol dot indicates derivative with respect
to the cosmic time.
\section{The Lambert $W$ Function}\label{sec-lambert}
We now turn our attention towards the Lambert $W$ function which holds a central place in the present work. The Lambert $W$ function, also sometimes referred to as the ``omega function" or the ``product logarithm", is defined mathematically as the multivalued inverse of the function $xe^{x}$, i.e.,
\begin{equation} \label{lw-1}
\text{Lambert}W(y) \cdot e^{\text{Lambert}W(y)}=y.
\end{equation} 
Eq. (\ref{lw-1}) has two real solutions if $-\frac{1}{e} \leq y < 0$, which correspond to two real branches of $\text{Lambert}W$\footnote{Note here that $W(y)$ at $y=-e^{-1},0,1$ can be computed as $-1,0,0.567143$ respectively. These three values might prove useful for our work.} \cite{Veberic1}. However, infinitely many solutions of Eq. (\ref{lw-1}) can be obtained with imaginary values of $y$ which shall correspond to infinitely many imaginary branches \cite{Corless1,Corless2}. Euler \cite{Euler1} is often credited with the earliest mention of Eq. (\ref{lw-1}) but Euler himself credited Lambert for his earlier work on the transcendental equation of the form \cite{Lambert1}
\begin{equation} \label{lte}
x^m-x^n=(m-n)\nu x^{m+n},
\end{equation}
where $m,n,\nu$ are constants. As a matter of fact, Lambert initially obtained a series solution in $p$ of the trinomial equation \cite{Corless1}
\begin{equation} \label{trineq}
x=p+x^{\alpha}
\tag{2'}
\end{equation}
and later extended the series to find powers of $x$ as well \cite{Lambert1,Lambert2}. Euler \cite{Euler1} used the substitution $x^{-n}$ for $x$ and setting $\alpha = mn$ and $p=(m-n)\nu$. to transform Eq. (\ref{trineq}) into the more symmetrical form given in Eq. (\ref{lte}).\\

We can compute the $n^{\text{th}}$ derivatives of the Lambert $W$\footnote{Henceforth, we shall write Lambert $W$ simply as $W$. Anyone interested in the history behind the choice of the letter $``W"$ is referred to the article by Hayes \cite{Hayes1}.} function as 
\begin{equation}
W^{n}(y)=\frac{W^{n-1}(y)}{y^n[1+W(y)]^{2n-1}}\varphi _{k=1}^{n} \delta _{kn}W^{k}(y),~~~~y \neq -\frac{1}{e},
\end{equation}
where $\delta _{kn}$ is the number triangle
\[ \begin{array}{ccccc}
\phantom{+}1 & & & &\\
-2 & -1 & & & \\
\phantom{+}9 & \phantom{+}8 & \phantom{+}2 & & \\
-64 & -79 & -36 & -6 & \\
\phantom{+}625 & \phantom{+}974 & \phantom{+}622 & \phantom{+}192 & \phantom{+}24  
\end{array} .\]
Particularly, the first order derivative of $W(y)$ can be evaluated as 
\begin{eqnarray}
W'(y) &=& \frac{W(y)}{y[1+W(y)]}, ~\mbox{if~} y \neq 0 \nonumber \\
&=& \frac{e^{-W(y)}}{1+W(y)}.
\end{eqnarray}
The antiderivative of $W(y)$ is given as
\begin{equation}
\int W(y)\text{d}y=y\left[W(y)-1+\frac{1}{W(y)}\right]+C,
\end{equation}
where $C$ is the arbitrary constant of integration.\\

We have defined and outlined the basic properties of the Lambert $W$ function in the last two paragraphs. Many other mathematical properties of this special function can be found in Refs. \cite{Veberic1,Mezo1,Cranmer1,http2}. It is remarkable to know that numerous real-life applications of the Lambert $W$ function can be found in Mathematics, Physics, and Computer Science. For an extensive discussion on some of such applications, one may see the article by Corless \cite{Corless1}. In General Relativity, the Lambert $W$ function is employed in finding solutions to the (1+1)-gravity problem \cite{Scott4} and in finding inverse of Regge-Finkelstein coordinates \cite{Regge1}.\\

Motivated by the above facts, in the next section, we explore its implications in studying the cosmic history of the Universe.

\section{Cosmology with the Lambert $W$ Function}\label{sec-model}
Saha and Bamba \cite{Saha1} recently studied the Lambert $W$ function in the context of Cosmology. They were the first to propose a novel equation of state (EoS) parameter which incorporates this function in a special fashion. It is worthwhile to mention here that the Lambert $W$ function appears while deriving solutions of the continuity equation in the gravitational particle creation scenario \cite{Chakraborty1}. This served as a motivation for them to study the evolutionary history of the Universe with the Lambert $W$ function. They assumed a spatially flat Friedmann-Robertson-Walker (FRW) universe as the spacetime metric and considered a perfect fluid having an effective EoS \cite{Saha1}
\begin{equation} \label{eos}
w_{\mathrm{eff}} = \frac{P}{\rho} = \left[\vartheta _\mathrm{1}\text{ln}\left\{W\left(\frac{a}{a_0}\right)\right\}+\vartheta _\mathrm{2}\left\{W\left(\frac{a}{a_0}\right)\right\}^3\right],
\end{equation}
where $P$ and $\rho$ are the pressure and energy density of the cosmic fluid, respectively. $a_{0}$ is the value of the scale factor, $a$, at the present epoch, while $\vartheta _\mathrm{1}$ and $\vartheta _\mathrm{2}$ are dimensionless parameters which must be fixed from observational data. The proposed EoS looks phenomenological and seems to be a bit speculative at first, but, theoretically, this EoS has been predicted to smoothly describe the evolutionary history of the Universe \cite{Saha1}. An important advantage of the form of EoS assumed in equation (\ref{eos}) is that it is independent of any prior assumption about the nature of dark energy. The expression for the energy density, $\rho$, can be obtained from the continuity equation as ($\rho_{0}$ is a positive constant) \cite{Saha1}
\begin{equation}
\rho = \rho _{0} \text{exp}\left[-3\left\{\text{ln}[W(a)][\vartheta _\mathrm{1} W(a)+\vartheta _\mathrm{1} +1]+W(a)(1-\vartheta _\mathrm{1})+\frac{\vartheta _\mathrm{2}}{12}W(a)^3[4+3W(a)]\right\}\right],
\end{equation}
while the deceleration parameter $q$, in terms of redshift $z$, is given by \cite{Saha1}
\begin{equation}
q=-\frac{\ddot{a}}{aH^2}=\frac{3}{2}\left\{1+\vartheta_\mathrm{1}\text{ln}\left[W\left(\frac{1}{1+z}\right)\right]+\vartheta _\mathrm{2}W\left(\frac{1}{1+z}\right)^3\right\}-1.
\end{equation}
where, $z=\frac{1}{a}-1$. Note that the functional form of $q(z)$ depends crucially on the values of the model parameters $\vartheta _\mathrm{1}$ and $\vartheta _\mathrm{2}$. In the next section, we first constrain these parameters using the observational data and with the best fit values obtained, we then try to reconstruct the functional dependence of $q(z)$. 
\section{Observational Constraints on the Lambert $W$ EoS}\label{sec-obs}
The cosmological model, discussed in the present context,
has been confronted with latest cosmological observations. The observational data, used to constraint the model parameters are briefly discussed in the following.\\
\par It is well known that the observational Hubble parameter dataset (OHD) is one of the most robust probes to analyze different cosmological models for its model independent nature. Recently, a plethora of papers have been published, for example, \cite{hp1,hp2,hp3,hp4,hp5,hp6,Magana:2018hcomp}, which determine the dynamical characteristics of many cosmological models. In this work, we consider the latest 51 data points of $H(z)$ measurements in the redshift range $0.07 \leq z \leq 2.36$, obtained from different surveys \cite{Zhang:2012mp,Stern:2009ep,Moresco:2012jh,Moresco:2016mzx,Gaztanaga:2008xz,Oka:2013cba,
Wang:2016wjr,Chuang:2012qt,Alam:2016hwk,Blake:2012pj,Ratsimbazafy:2017vga,
Anderson:2013oza,Moresco:2015cya,Bautista:2017zgn,Delubac:2014aqe,Font-Ribera:2013wce} and the corresponding $H(z)$ values are given in  \cite{Magana:2018hcomp}. Among them 31 data points calculated from the  differential age method (i.e., cosmic chronometers technique), however, 20 data points of this sample are calculated from the BAO measurements under different fiducial cosmologies based on the standard $\Lambda$CDM model. Although some of the data points from the BAO measurements are being correlated, however, we assume here that they are independent measurements. On the other hand, the cosmic chronometers method \cite{jimccm2002} offers to directly measure the expansion rate of the
universe (i.e., $H(z)$) using spectroscopic dating of passively-evolving galaxy to compare their ages, providing $H(z)$ measurements that are model-independent. Note that these data points (31 points) constitute the majority of our $H(z)$ sample. Additionally, the latest SH$0$ES measurement of the Hubble constant $H_{0} = 74.03 \pm 1.42$ km/s/Mpc (at 68\% CL) \cite{r19}, denoted as R19, is also included in the analysis. \\

We use the above sample to constrain the free parameters of the model as given in equation (\ref{eos}), and search for an alternative solution to the accelerated expansion of the Universe. The $\chi^{2}$ function for this dataset is defined as 
\begin{equation}\label{eqchi2h}
\chi^2= \sum^{N}_{i=1}\frac{[{H}_{obs}(z_{i}) - {H}_{th}(z_{i},\theta_p)]^2}{\sigma^2_{H}(z_{i})} 
\end{equation}
where $N$ stands for the number of the observational Hubble parameter ${H}_{obs}(z_{i})$ at $z_{i}$ and $\sigma_{H}(z_{i})$ represents the error associated with the $i^{th}$ data point.  Also, ${H}_{th}(z_{i}, \theta_p)$ stands for the
theoretical Hubble parameter for a given model depending on model parameters $\theta_1$, $\theta_2$ ... $\theta_p$. One can now use the maximum likelihood method and take the likelihood
function as
\begin{equation}
{\cal L}={\rm exp}\left[-\frac{\chi^2}{2}\right]
\end{equation}
The best-fit corresponds to the free parameters for which $\chi^2$
function is minimized (say, $\chi^{2}_{min}$). In this work, we have minimized
$\chi^{2}$ with respect to the parameters $\vartheta _\mathrm{1}$ and $\vartheta _\mathrm{2}$ to calculate their best-fit values. In what follows, we discuss the results obtained from the statistical analysis of the above mentioned datasets.\\
\par
\begin{figure}[htp]
\begin{center}
\includegraphics[width=9.5 cm]{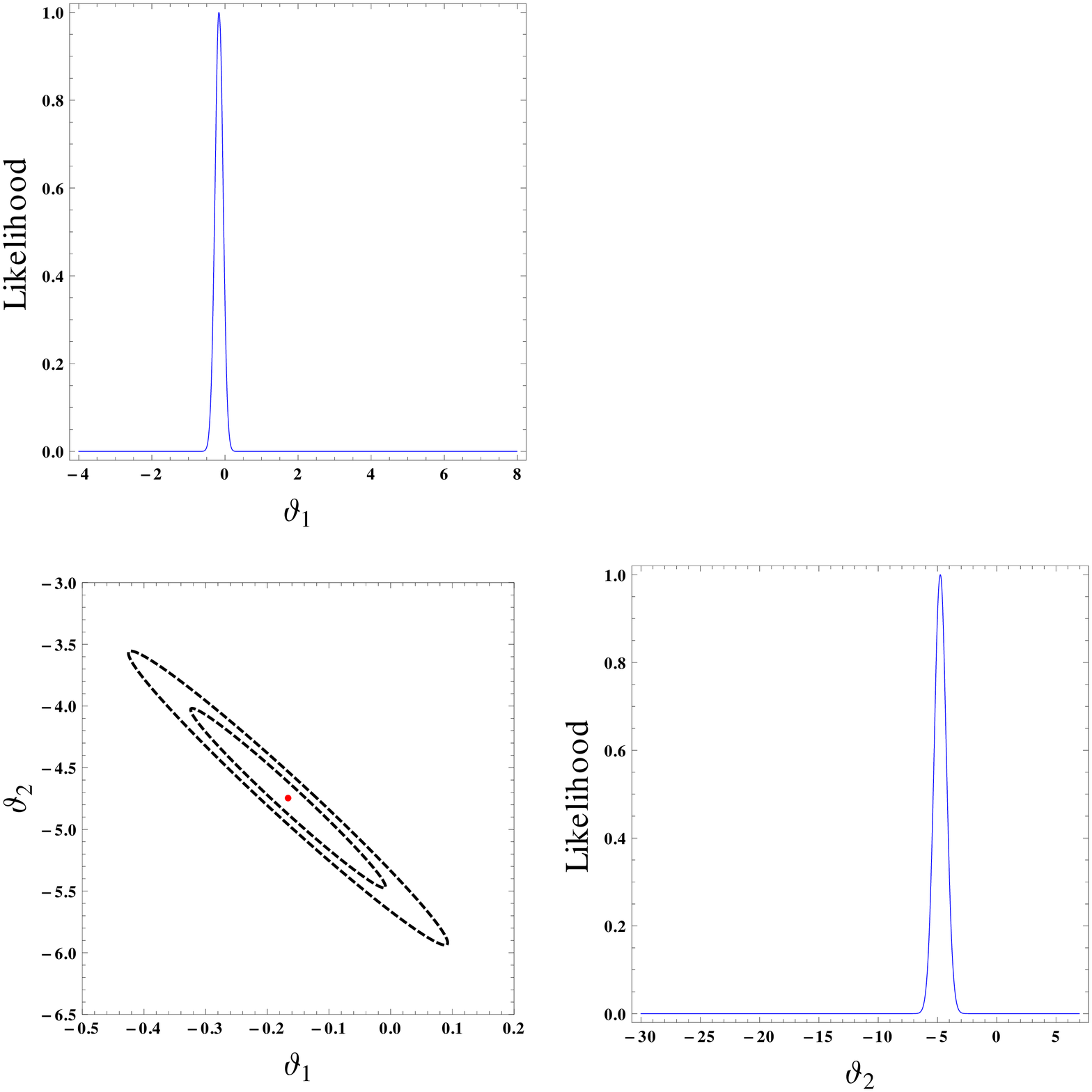}
\caption{Marginalized posterior distribution of the set of parameters ($\vartheta _\mathrm{1}$, $\vartheta _\mathrm{2}$) and corresponding 2D confidence contours obtained from the $\chi^{2}$ analysis for the present model utilizing the joint OHD+R19 dataset. The red point represents the best-fit values of the parameter pair ($\vartheta_{1},\vartheta_{2}$).}\label{figc}
\end{center}
\end{figure}

\begin{figure}[htp]
\begin{center}
\includegraphics[width=7 cm]{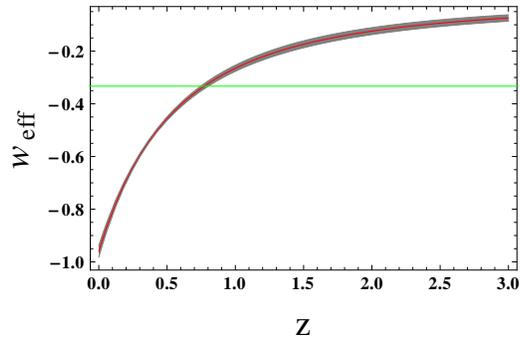}
\caption{Reconstructed $w_{eff}$ as a function of redshift $z$. In this plot, the red curve corresponds the evolution of $q(z)$ for the best-fit case and the gray shaded region indicates $1\sigma$ error region. Here, the horizontal green line is for $w_{eff}=-\frac{1}{3}$.}\label{figweff}
\end{center}
\end{figure}

\begin{figure}[htp]
\begin{center}
\includegraphics[width=7 cm]{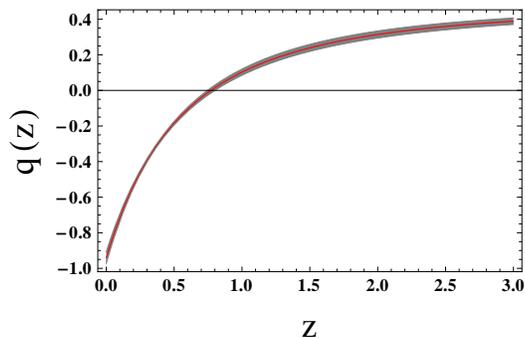}
\caption{Plot of the deceleration parameter $q$ as a function of redshift $z$ is shown in $1\sigma$ error region (gray) by considering OHD+R19 dataset. Here,
the red curve represents the corresponding evolution of $q$ for the best-fit case and the horizontal line stands for $q(z)=0$.}\label{figq}
\end{center}
\end{figure}

\begin{figure}[htp]
\begin{center}
\includegraphics[width=6.5 cm]{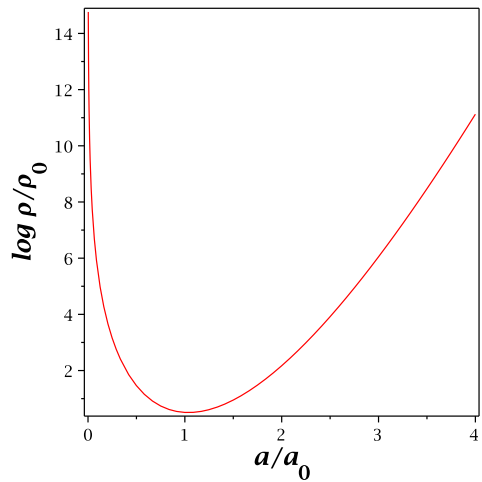}
\caption{This figure shows the evolution of the normalized energy density $\rho/\rho_0$ in the logarithmic scale against the normalized scale factor $a/a_0$. We have assumed $a_0=1$ without any loss of generality and considered the best-fit values, $\vartheta _\mathrm{1}=-0.166$ $\&$ $\vartheta _\mathrm{2}=-4.746$, of the coefficients in the Lambert $W$ EoS parameter.}\label{figrho}
\end{center}
\end{figure}

\begin{figure}[htp]
\begin{center}
\includegraphics[width=6.5 cm]{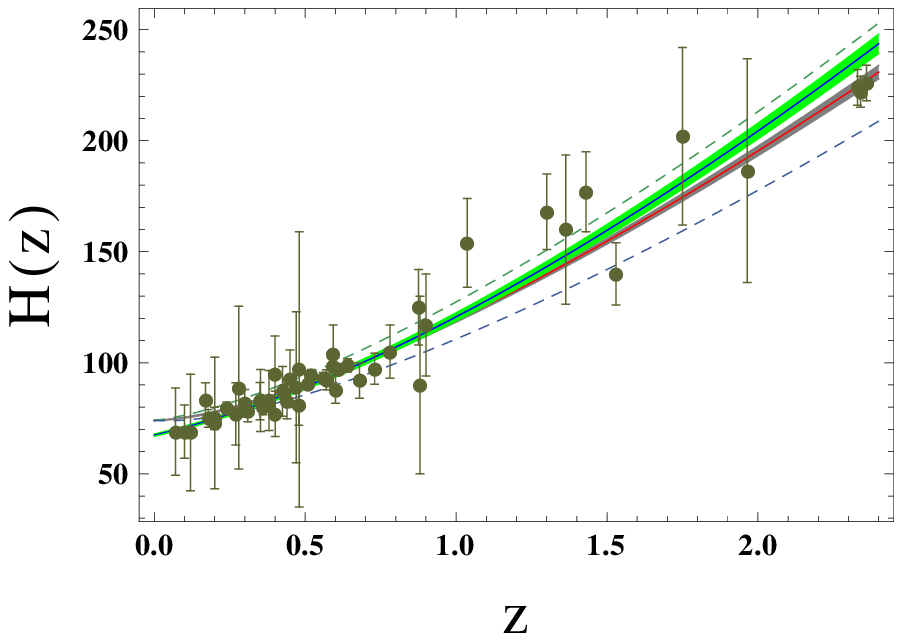}\hspace{5mm}\includegraphics[width=6.8cm]{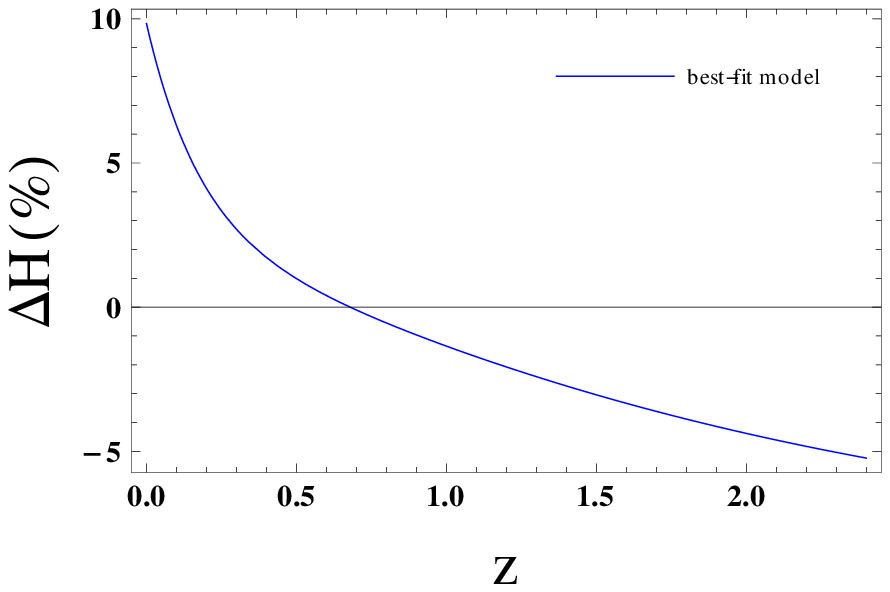}
\caption{Left panel: The evolution of $H(z)$ within $1\sigma$ (gray) and $2\sigma$ (dashed) confidence levels are shown for the present model by considering the
OHD+R19 dataset. In this plot, the dots correspond to the 51 $H(z)$ data points, whereas the green shaded contour ($1\sigma$) indicates the corresponding evolution of $H(z)$ for a flat $\Lambda$CDM model \cite{pl19}. For each model, the solid curve inside the shaded region corresponds the evolution of $H(z)$ for the best-fit case. Right Panel: The corresponding relative difference $\bigtriangleup H(\%)=100\times [H_{{\rm Lambert}W}(z)-H_{\Lambda CDM}(z)]/H_{\Lambda CDM}(z)$. }\label{figH}
\end{center}
\end{figure}

\begin{figure}[htp]
\begin{center}
\includegraphics[width=6.5 cm]{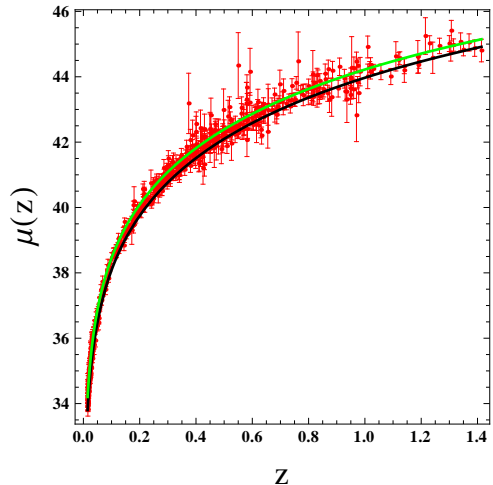}
\caption{This figure shows the Error bar plot of 580 points of Union 2.1 compilation Supernovae Type Ia data sets (red dots) \cite{snia} together with the presented model shown in solid black line with $\vartheta _\mathrm{1}=-0.166$ $\&$ $\vartheta _\mathrm{2}=-4.746$. The standard $\Lambda$CDM model is also shown in solid green line for model comparision. Here, $\mu(z)$ denotes distance modulus, which is the
difference between the apparent magnitude and the absolute magnitude of the observed supernova, is given by \cite{Copeland1} $\mu(z)=5{\rm log}_{10}(\frac{d_{L}}{\rm Mpc})+25$, where $d_{L}$ is the luminosity distance.}\label{figmu}
\end{center}
\end{figure}

 {\textbf{Results:}} Figure \ref{figc} shows the $1\sigma$ and $2\sigma$ confidence level confidence contours on the set of parameters ($\vartheta _\mathrm{1}$, $\vartheta _\mathrm{2}$) and the marginalized likelihood function of the present model obtained in the combined analysis with the combinations of the datasets OHD and R19. The best-fit values for the model parameters are obtained as $\vartheta _\mathrm{1}=-0.166\pm 0.104$ ($1\sigma$) and $\vartheta _\mathrm{2}=-4.746\pm 0.479$ ($1\sigma$) with $\chi^{2}_{min}=36.853$.
Figure \ref{figweff} shows the evolution of the effective EoS parameter $w_{eff}(z)$ within $1\sigma$ error region for our model. The reconstruction of $w_{eff}(z)$ has been done by the joint (OHD+R19) dataset. It has been observed from figure \ref{figweff} that $w_{eff}(z)$ tends to zero at high redshift for any values of $\vartheta_{1}$ and $\vartheta_{2}$ and thus it become indistinguishable from the dark matter component at high redshift. On the other hand, $w_{eff}(z)$ enters in the quintessence regime ($-1 < w_{eff} <-\frac{1}{3}$, within $1\sigma$ confidence level) at relatively low redshifts and its present value is found to be $-0.96\pm0.02$ ($1\sigma$). Therefore, the functional form of $w_{eff}(z)$, as given in equation (\ref{eos}), can easily accommodate both the phases of the Universe, i.e., early matter dominated era and late-time dark energy dominated era. It is also evident that the present value of $w_{eff}$ is very close to the Cosmological Constant $\Lambda$. This implies that our proposed EoS might serve as a unification of dark matter and dark energy.
Similarly, the evolution of the deceleration parameter $q(z)$ within $1\sigma$ confidence level is shown in figure \ref{figq}. The redshift at which $q$ changes sign from positive to negative corresponds to the onset of late-time cosmic acceleration. The redshift around which the transition from the decelerating ($q>0$) expansion to the accelerating ($q<0$) expansion occurs is found to be $0.77\pm0.03$ ($1\sigma$). The results are in good agreement with the measured transition redshift $z_{t}$ based on the OHD (cosmic chronometer) dataset \cite{hp2,hp3,hp4} including the standard $\Lambda$CDM prediction ($z_{t}\approx 0.7$). On the other hand, in figure \ref{figrho}, we have plotted the evolution of the normalized energy density $\rho/\rho_0$ in the logarithmic scale against the normalized scale factor $a/a_0$. We have assumed $a_0=1$ without any loss of generality and considered the best-fit values, $\vartheta _\mathrm{1}=-0.166$ $\&$ $\vartheta _\mathrm{2}=-4.746$, of the coefficients in the Lambert $W$ EoS parameter. The figure clearly shows that the Lambert $W$ EoS exhibits a transition of a dark matter dominated era to a dark energy dominated era with the evolution of the Universe. Interestingly enough, though, it will allow the Universe to transit from the dark energy era to a future dark matter dominated era. Furthermore, in the left panel of figure \ref{figH}, we have shown the evolution of the Hubble parameter $H(z)$ within $2\sigma$ confidence level for our model and have compared that with the latest 51 points of $H(z)$ dataset \cite{Magana:2018hcomp} as well as the flat $\Lambda$CDM model. From this figure, we have observed that the model is well consistent with the OHD+R19 dataset against redshift parameter. In the right panel of figure \ref{figH}, we observed that for the best-fit case, the relative difference $\bigtriangleup E$ is close to 0.98$\%$  at $z\sim 0.5$, while the differences between the present and $\Lambda$CDM models are negligible around $z\sim 0.67$. It has also been observed that $H_{{\rm Lambert}W}(z)<H_{\Lambda CDM}(z)$ at high redshifts, whereas  $H_{{\rm Lambert}W}(z)>H_{\Lambda CDM}(z)$ at relatively low redshifts. Finally, the best fit of distance modulus $\mu(z)$ as a function of $z$ for the present model and the 580 points of Union 2.1 compilation \cite{snia} Supernovae Type Ia (SNIa) datasets  are plotted in figure \ref{figmu}. The evolution of $\mu(z)$ for the standard $\Lambda$CDM model is also shown in figure \ref{figmu} for model comparision. From figure \ref{figmu}, we have found that the Lambert $W$ model reproduces the observed values of $\mu(z)$ quite effectively. Furthermore, we have checked that the nature of the evolution of $\mu(z)$ is hardly affected by a small change in the values of $\vartheta _\mathrm{1}$ and $\vartheta _\mathrm{2}$ within $1\sigma$ confidence limit.
\section{Conclusions}\label{sec-con}
In summary, we have investigated the possibility that the Universe is driven by a single dark fluid described by a Lambert $W$ EoS parameter which is dependent on two free parameters $\vartheta_{1}$ and $\vartheta_{2}$. We have then fixed the values of $\vartheta_{1}$ and $\vartheta_{2}$ from the analysis of recent observational datasets. As discussed in the previous section, the measurements of Hubble parameter at different redshift from the differential age of galaxies and the BAO methods are incorporated in the present analysis. Also, the latest measurement of $H_{0}$ from \cite{r19} is also taken into account. The best fit values of ($\vartheta _\mathrm{1}$, $\vartheta _\mathrm{2}$) for the combined OHD+R19 dataset are obtained as $\vartheta _\mathrm{1}=-0.166$ and $\vartheta _\mathrm{2}=-4.746$. Furthermore, using the best fit values of $\vartheta _\mathrm{1}$ and $\vartheta _\mathrm{2}$, we have reconstructed the evolutions of $w_{eff}(z)$, $q(z)$ and $H(z)$ for the present model. The main results of our study are summarized as
follows.\\
\par We have found that the effective EoS parameter $w_{eff}(z)$ can easily accommodate both the phases of cosmic evolution, i.e., early matter (dust) dominated phase and late-time dark energy dominated phase. Additionally, we have observed that $w_{eff}$ remains in the quintessence regime. Its present value has been found to be $-0.96\pm0.02$ ($1\sigma$). This shows that the present value of $w_{eff}$ is very close to the cosmological constant $\Lambda$. Thus, our proposed EoS parameter might serve as a unification of dark matter and dark energy. It has also been found that the deceleration parameter $q$ undergoes a smooth transition from a decelerated ($q>0$) to an accelerated ($q<0$) phase of expansion at the redshift $z_{t}=0.77\pm0.03$ ($1\sigma$). This result is in good agreement with the measured $z_{t}$ based on the cosmic chronometer dataset \cite{hp2,hp3,hp4} including the standard $\Lambda$CDM prediction ($z_{t}\approx 0.7$). It is worth emphasizing that the evolution scenarios of $w_{eff}(z)$ and $q(z)$ are necessary to explain both the observed growth of structures at the early epoch and the late-time cosmic acceleration measurements. Finally, we have shown the evolution of $H(z)$ within $2\sigma$ confidence level for our model and have compared that with the latest 51 points of $H(z)$ dataset \cite{Magana:2018hcomp} as well as the $\Lambda$CDM model. For the best-fit case, we observed that the relative difference $\bigtriangleup E$ is close to 0.98$\%$  at $z\sim 0.5$, while the differences between the two models are negligible around $z\sim 0.67$. We have also found that the present model reproduces the observed values of the distance modulus $\mu(z)$ quite effectively (see figure \ref{figmu}). In this context, it desreves to mention here that in a recent work \cite{growthus}, a group of authors including us, have studied the effect on the growth of perturbations for the Lambert $W$ dark energy model. We have performed the analysis for two different cosmological scenarios. In the first case, we have considered the universe to be filled with two different fluid components, namely, the LambertW dark energy component and the baryonic matter component, while in the second case, we have considered that there is a single fluid component in the universe whose EoS parameter is described by the Lambert $W$ function. We have then compared the growth rates of the Lambert $W$ model with that for a $\Lambda$CDM model as well as the CPL model. It has been observed that the presence of Lambert $W$ dynamical dark energy sector changes the growth rate and affects the matter fluctuations in the Universe to a great extent. \\ 
\par We conclude that the Lambert $W$ EoS parameter provides some interesting consequences in the cosmological perspective, and thus it can be a candidate for the description of nature. However, it is natural to extend the present work with addition of other datasets from Supernovae Type Ia, BAO and CMB probes in order to constrain the new parameters $\vartheta _\mathrm{1}$ and $\vartheta _\mathrm{2}$ more precisely. The present analysis is one preliminary step towards that direction.

\end{document}